# Comparing an Ensemble Kalman Filter to a 4DVAR Data Assimilation System in Chaotic Dynamics

Fabrício Pereira Harter[1], Cleber Souza Corrêa[2]

**ABSTRACT:** In this paper, the Ensemble Kalman Filter is compared with a 4DVAR Data Assimilation System in chaotic dynamics. The Lorenz model is chosen for its simplicity in structure and the dynamic similarities with primitive equations models, such as modern numerical weather forecasting. It was examined if the Ensemble Kalman Filter and 4DVAR are effective to track the Control for 10, 20 and 40% of error at the Initial Conditions. With 10% of noise, the trajectories of both are almost perfect. With 20% of noise, the differences between the simulated trajectories and the observations as well as "true trajectories" are rather small for the Ensemble Kalman Filter but almost perfect for 4DVAR. However, the differences are increasingly significant at the later part of the integration period for the Ensemble Kalman Filter, due the chaotic behavior system. However, for the case with 40% error at the Initial Condition, neither the Ensemble Kalman Filter or 4DVAR could track the Control with only 3 observations ingested. To evaluate a more realistic assimilation application, it was created an experiment in which the Ensemble Kalman Filter ingested single observation at the 180[th] time step in the $X$, $Y$, and $Z$ Lorenz variables and only in the $X$ variable. The results show a perfect fit of 4DVAR and the Control during a complete integrations period, but the Ensemble Kalman Filter has a disagreement after the 80[th] time step. On the other hand, it was shown a considerable disagreement between the Ensemble Kalman Filter trajectories and the Control as well as a total fail of 4DVAR. Better results were obtained for the case in which observation covers all the components of the model vector.

# INTRODUCTION

Numerical models are an important tool to weather forecasting, probably the most of all, being largely applied to specific issues, such as aeronautic weather forecasting. The development to improve these models is an ongoing process, and Data Assimilation has been a significant research line to improve weather forecasting model and its applications, such as wind profile predictive models, applied by the Brazilian Air Force at the Centro de Lançamento de Alcântara (CLA), Maranhão State.

Data Assimilation is a procedure to get the Initial Condition as accurately as possible, through the statistical combination of collected observations and a background field, usually a short-range forecast; such a best estimate is called "analysis" in Meteorology. The Data Assimilation community has extensively tested 2 approaches, one based on Kalman filtering and another, on variational calculation.

Several operational meteorological centers, such as the European Centre for Medium-Range Weather Forecasts (ECMWF), used to have Three-Dimensional Data Assimilations (3DVAR) to build their analysis (Parrish and Derber 1992; Courtier *et al.* 1998). 3DVAR is a kind of accurate statistical interpolation, being computationally feasible. However, it does not consider the "errors of the day". It means that the background error covariance is static. To overcome this failure, 4DVAR (Courtier and Talagrand 1998; Rabier and Courtier 1992; Courtier *et al.* 1994; Rabier *et al.* 2000) and Kalman filters (Miller *et al.* 1994) have been subject of many researches.

1. Universidade Federal de Pelotas – Faculdade de Meteorologia – Departamento de Meteorologia – Pelotas/RS – Brazil. 2. Departamento de Ciência e Tecnologia Aeroespacial – Instituto de Aeronáutica e Espaço – Divisão de Ciências Atmosféricas – São José dos Campos/SP – Brazil.

Author for correspondence: Fabrício Pereira Harter | Universidade Federal de Pelotas – Faculdade de Meteorologia – Departamento de Meteorologia | Rua Gomes Carneiro, 1 – Centro | CEP: 96.010-610 – Pelotas/RS – Brazil | Email: fabricio.harter@ufpel.edu.br

The 4DVAR is model-dependent and computationally much more expensive than 3DVAR, but more accurate. In turn, only reduced rank Kalman Filters (KF) can be applied to operational numerical weather forecasting owing the computational cost. KF are especially interesting due the capacity to estimate the error covariance of the analysis.

The large dimension of the models can prohibit the implementation of KF in operational numerical weather prediction, but its implementation with simplifications — such as Ensemble Kalman Filter (EnKF; Evensen 1994; Burgers *et al.* 1998; Whitaker and Hamill 2002; Tippet *et al.* 2003), Ensemble Transform Kalman Filter (ETKF; Bishop *et al.* 2001), Ensemble Adjustment Kalman Filter (EAKF; Anderson 2001) and Local Ensemble Kalman Filter (LEKF; Ott *et al.* 2002, 2004) — is possible due to the decrease in the computational cost.

The purpose of this research was to compare an EnKF to a 4DVAR Data Assimilation system through the Lorenz equations and the sensitivity of the model to Initial Condition (IC) for non-linear and chaotic regimes. The main objective was to compare an implementation of the EnKF to a 4DVAR method with different noise levels at the IC and explore the assimilation systems over-determined by lack of observation.

The organization of the paper is as follows: the "Methodology" section describes a brief theoretical formulation of Lorenz equations and presents a basic introduction to EnKF and 4DVAR; numerical experiments are shown in the "Results" section and final comments are found in the "Final Comments" section.

## METHODOLOGY
### LORENZ EQUATIONS

Lorenz was looking for the periodic solutions of the Saltzman's model (1962), considering a spectral Fourier decomposition and only low-order terms. Then, he obtained the following system of non-linear coupled ordinary differential equations:

$$dX/d\tau = -\sigma(X - Y) \quad (1)$$

$$dY/d\tau = rX - Y - XZ \quad (2)$$

$$dZ/d\tau = XY - bZ \quad (3)$$

where $\tau \equiv \pi^2 H^{-2}(1+a^2)\kappa t$ is the non-dimensional time, being $H$, $a$, $\kappa$ and $t$ layer height, thermal conductivity, wave number (diameter of the Rayleigh-Bénard cell) and time, respectively; $\sigma \equiv \kappa^{-1}\nu$ is the Prandtl number, being $\nu$ the kinematic viscosity; $b \equiv 4(1+a^2)^{-1}$ is the relation between the height and the width of the rectangle (orbit travelled; Saltzman 1962); the parameter $r = R/R_c \propto \Delta T$ is the Rayleigh number, being $T$ the temperature and $R_c$ the critical Rayleigh number.

### ENSEMBLE KALMAN FILTER

Kalman Filter is the best linear unbiased estimator for a linear model under Gaussian assumption for the measurements and model random errors. The Kalman Filter method for non-linear models is called Extended Kalman Filter (EKF) and is given by the following definition, considering forecast and analysis procedures:

*Forecast*

$$w^f_{n+1} = F_n w^f_n + \mu_n \quad (4)$$

$$P^f_{n+1} = F_n P^a_n F^T_n + Q \quad (5)$$

*Analysis*

$$K_{n+1} = P^f_{n+1} H^T_{n+1} [R_{n+1} + H_{n+1} P^f_{n+1} H^T_{n+1}]^{-1} \quad (6)$$

$$w^a_{n+1} = w^f_{n+1} + K_{n+1}\left[y^o_{n+1} - H(w^f_{n+1})\right] \quad (7)$$

$$P^a_{n+1} = [I - K_{n+1} H_{n+1}] P^f_{n+1}, \quad (8)$$

where $F_n$ is our mathematical model; $\mu_n$ is the stochastic forcing (modeling noise error); subscript $n$ denotes discrete time-step; superscript $f$ represents the forecasting value. The observation system $[\overset{\circ}{y}_{n+1} - H(w^f_{n+1}) + v_n]$ is modeled by the matrix $H$, and $v_n$ is the noise associated to the $Y$ observation. The typical gaussianity, 0-mean and ortogonality hypotheses for the noises are adopted. The state vector is defined as $w_{n+1} = [X_{n+1}, Y_{n+1}, Z_{n+1}]$ and estimated through the recursion $w^a_{n+1} = w^f_{n+1} + K_{n+1}[\overset{\circ}{y}_{n+1} - H(w^f_{n+1})]$, where $w^a_{n+1}$ is the analysis value, $K_n$ is the Kalman gain, computed from the minimization of the estimation error variance $J_{n+1}$ (Lorenz 1963), being $J_{n+1} = E\{(w^a_{n+1} - w^f_{n+1})^T(w^a_{n+1} - w^f_{n+1})\}$, being $E\{.\}$ the expected value; $Q$ is the covariance of $\mu_n$ and $R_n$ is the covariance of $v_n$. The assimilation is done through the sampling: $r_{n+1} \equiv y_{n+1} - y^f_{n+1} = y_{n+1} - H_n w^f_{n+1}$; $P^a$ and $P^f$ are forecast and analysis covariance error, respectively.

According to Kalnay (2004), "Even if a system starts with a poor initial guess of the state of the atmosphere, the EKF may go through an initial transient period of a week or so, after which it should provide the best unbiased estimate of the state of the atmosphere and its error covariance". However, according to Miller *et al.* (1994), if the system is very unstable and the observation is not frequent enough, it is possible for the linearization to became inaccurate, and the EKF may drift away from the true solution.

The updating of Eq. 5 provides the "errors of the day", but its computational cost makes this updating impossible in practice to carry out. Therefore, this equation was replaced by simplifying assumptions, such as ensemble mean. In this study, it was proposed an EnKF which consists of replacing the forecast error covariance (Eq. 5) by:

$$P^f \approx \frac{1}{K-1} \sum_{k=1}^{K} \left( X_k^f - \bar{X}^f \right)\left( X_k^f - \bar{X}^f \right)^T \quad (9)$$

where the ensemble has $K$ data assimilation cycles; $k$ is the iteration.

## FOUR-DIMENSIONAL VARIATIONAL DATA ASSIMILATION

In order to formulate the 4DVAR, first of all, a cost function $J$ should be introduced to measure the misfit between the model and observations:

$$J = \frac{1}{2} \sum_{k=0}^{k=kmax} (w_k^f - w_k^0)^T R_n^{-1} (w_k^f - w_k^0) \quad (10)$$

In this approach, the model is used as a strong constraint (Sasaki 1970); it means that the model is considered perfect in the cost function formulation.

To minimize the difference between the model state and the observation, it could be used an iterative process where the values of the initial ones and model parameters are adjusted in the opposite direction of the gradient of the cost function. In operational weather forecasting, efficient minimizations methods are necessary, such as quasi-Newton and conjugate-gradient. In this research, it was used a simple steepest descent approach with **α**, due the lower dimension of the Lorenz equations, given by:

$$w_0^{f,k+1} = w_0^{f,k+1} - \alpha \nabla J\left(w_0^{f,k}\right), \quad (11)$$

where $\alpha$ is a parameter to be chosen to achieve convergence through iterations; $\nabla J$ is the gradient of the cost function with respect to initial state $w_0^{f,k}$.

If the iteration Eq. 11 converges, $w_0^{f,k}$ will approximate the desired initial state $w_0^{f,\infty}$ that satisfies $J = \min (J)$. However, for operational primitive equation models, where the number of degrees of freedom is of the order of $10^7$, this approach has some limitations. One of the limitations is overcoming by the adjoint model integrated backwards in time, which achieves an exact cost function gradient. This methodology, called 4DVAR, is summarized in the next section.

## TANGENT LINEAR OPERATOR AND TANGENT LINEAR MODEL

Considering the model state $w_0^f$ and its small perturbation $w_0^{f,tl}$ (where $tl$ means tangent linear), the change in the cost function $J(w_0^f)$, caused by the small perturbation, is denoted by $J^{tl}(w_0^f)$, where:

$$J^{tl}\left(w_0^f\right) = J\left(w_0^f + w_0^{f,tl}\right) - J\left(w_0^f\right) \quad (12)$$

In the limit of $\| w_0^{f,tl} \| \to 0$, Eq. 12 becomes

$$J^{tl}\left(w_0^f\right) = [\nabla J(w_0^f)]^T w_0^{f,tl} \quad (13)$$

Using the definition of $J(w_0^f)$ in Eq. 10, Eq. 12 can be written as

$$J^{tl}\left(w_0^f\right) = \sum_k (w_k^f - w_k^o)^T\, w_k^{f,tl} \quad (14)$$

Combining Eqs. 13 and 14, one has

$$[\nabla J(w_0^f)]^T w_0^{f,tl} = \sum_k (w_k^f - w_k^o)^T\, w_k^{f,tl} \quad (15)$$

Equation 15 may be used to compute the gradient of the cost function $J(w_0^f)$, since the relationship between $w_k^{f,tl}$ and $w_0^{f,tl}$ is found. To find this relationship, linearization of the model $w_k^f = N(w_{k-1}^f)$ around the basic state, is done by:

$$w_{k+1}^{f,tl} = \frac{\partial N(w_k^f)}{\partial w^f} w_k^{f,tl} = L\left(w_k^f\right) w_k^{f,tl} \quad (16)$$

where $L(w_k^f)$ is the tangent linear operator which depends on the basic state $w_k^f$ and time step $k$.

Equation 16 is the tangent linear equation of the forward model (Eq. 4). Through this equation, iteratively, the desired relation between $w_k^{f,tl}$ and $w_0^{f,tl}$ is obtained by:

$$w_k^{f,tl} = L_k w_0^{f,tl} \qquad (17)$$

where:

$$L_k = L(w_{k-1}^f)L(w_{k-2}^f)\cdots L(w_1^f)L(w_0^f) \qquad (18)$$

Substituting Eq. 17 in Eq. 15, it is obtained

$$\nabla J(w_0^f) = \sum_k L_k^T(w_k^f - w_k^o) \qquad (19)$$

where $L_K^T$ is the transpose of $L_K$.

$$L_k^T = L^T(w_0^f)L^T(w_1^f)\cdots L^T(w_{k-2}^f)L^T(w_{k-1}^f) \qquad (20)$$

The transpose of the tangent linear operator is known as the adjoint operator. It is important to note that the order of the time index of the adjoint operator is reversed, compared to tangent linear operator. By definition, the adjoint model is:

$$w_{k-1}^{f,ad} = L_k^T(w_{k-1}^f)w_k^{f,ad} \qquad (21)$$

Thus, 4DVAR can be summarized by the procedures described below:

- Integration of forecast model (Eqs. 1 – 3) forward in time and storage of the trajectories.
- Integration of adjoint model backward in time (Eq. 21). During this integration, the observation is ingested whenever it exists. By this step, the cost function is minimized (Eq. 11).
- A new initial state is provided by Eq. 11.
- This new initial state is used to start the next $k$ cycle.

## RESULTS

### NUMERICAL EXPERIMENTS

In the results explored forward from here, the Lorenz equations are solved by finite differences with a non-dimensional time increment of 0.01 for 200 time steps integration length. According to Lorenz (1963), at a given $\sigma = 10$ and $b = 8/3$, the corresponding Rayleigh number is 24.74, which means that $r$ larger than 24.74 will make a chaotic system. In this paper, only the variable $X$ is depicted, just because $Y$ and $Z$ take to the same conclusions.

In this study, the sensitivity of the model (Classic Experiment) to the IC is evaluated. In Fig. 1a, it is depicted the resulting model state trajectories for $\sigma = 10$, $b = 8/3$ and $r = 10$, assuming the IC to be $X_0 = 1.00$, $Y_0 = 3.00$, $Z_0 = 5.00$ (Case 1) and assuming IC to be $X_0 = 1.10$, $Y_0 = 3.30$, $Z_0 = 5.50$ (Case 2). It means that Case 1 differs from Case 2 with an offset of 10% of noise for all model variables. In Fig. 1b, this experiment is directed to the chaotic regime ($r = 32$), Case 3 and Case 4.

This simple experiment shows that, for non-linear regime ($r = 10$), the noise at the IC is not determinant for the success of the prediction. By other side, for chaotic regime ($r = 32$), comparing the trajectories of Case 1 and Case 2, it is clearly seen that the bifurcation in the model solution occurs throughout the integration period. This experiment illustrates the sensitivity of the chaotic system to IC.

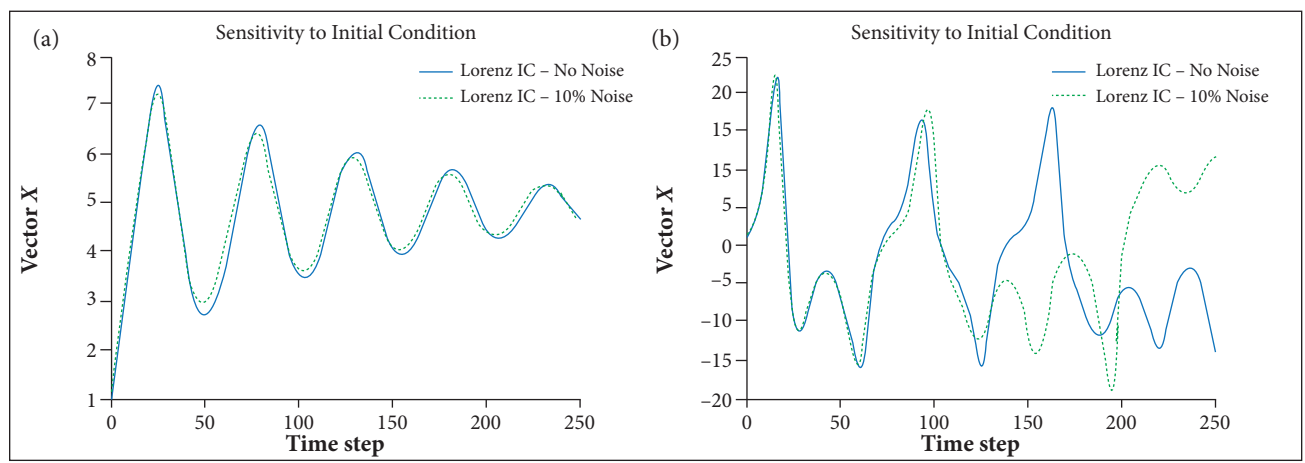

**Figure 1.** Resulting model state trajectories.

*Experiment 1*

In this experiment (Fig. 2), EnKF and 4DVAR assimilate the same evolution trajectories as the Classical Experiment, but the initial guess is rather poor: 10% ($X = 1.10$, $Y = 3.30$, $Z = 5.50$) — Fig. 2a; 20% ($X_0 = 1.20$, $Y_0 = 3.60$, $Z_0 = 6.00$) — Fig. 2b; and 40% ($X_0 = 1.40$, $Y_0 = 4.20$, $Z_0 = 7.00$) — Fig. 2c. It can be observed from the plots in Fig. 2 that, with 10% of noise, the trajectories of both EnKF and 4DVAR are almost perfect; with 20% of noise, the differences between the simulated trajectories and the observations as well as "true trajectories" are rather small for EnKF but almost perfect for 4DVAR. However, the differences are increasingly significant at the later part of the integration period for EnKF, due to the chaotic behavior of the system. However, for the case with 40% error at the IC, neither EnKF or 4DVAR could track the Control with only 3 observations ingested. In fact, according to the literature, given a 40% error at the IC, the chaotic nature of the model solution demands a relatively large number of observations in the EnKF assimilation and large data assimilation window in 4DVAR.

*Experiment 2*

Numerical Weather Forecasting is of the order of $16^6 – 10^7$ degrees of freedom, whereas the total number of conventional observations of the variables used in the models is of the order of $10^4$ (Kalnay 2004). Thus, the amount of observed variables applied hitherto is somewhat overestimated. By this reason, Experiment 2 represents a more realist assimilation application, where the EnKF ingests a single observation at the $180^{th}$ time step in the $X$, $Y$, and $Z$ Lorenz variables (Fig. 3a) and only in the *X* variable (Fig. 3b). In Fig. 3a, the results show a perfect fit of 4DVAR and the Control during a complete integrations period, but EnKF has a disagreement after the 80$^{th}$ time step. On the other hand, Fig. 3b shows a considerable disagreement between the EnKF trajectories and the Control and a total fail of 4DVAR. It is clear that better results are obtained for the case in which observation covers all the components of the model vector.

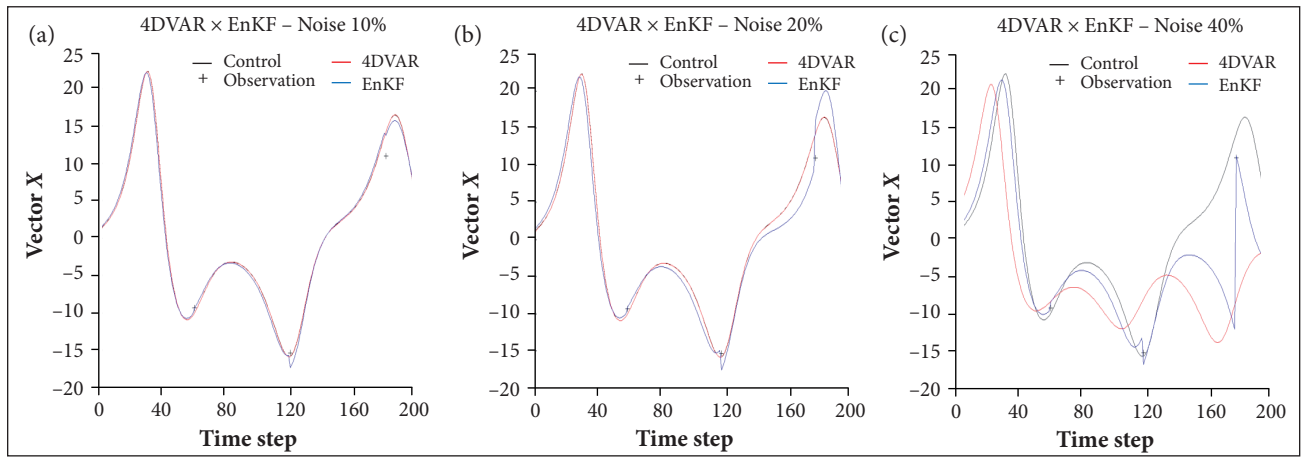

**Figure 2.** Evolution of the EnKF and 4DVAR for poor Initial Conditions.

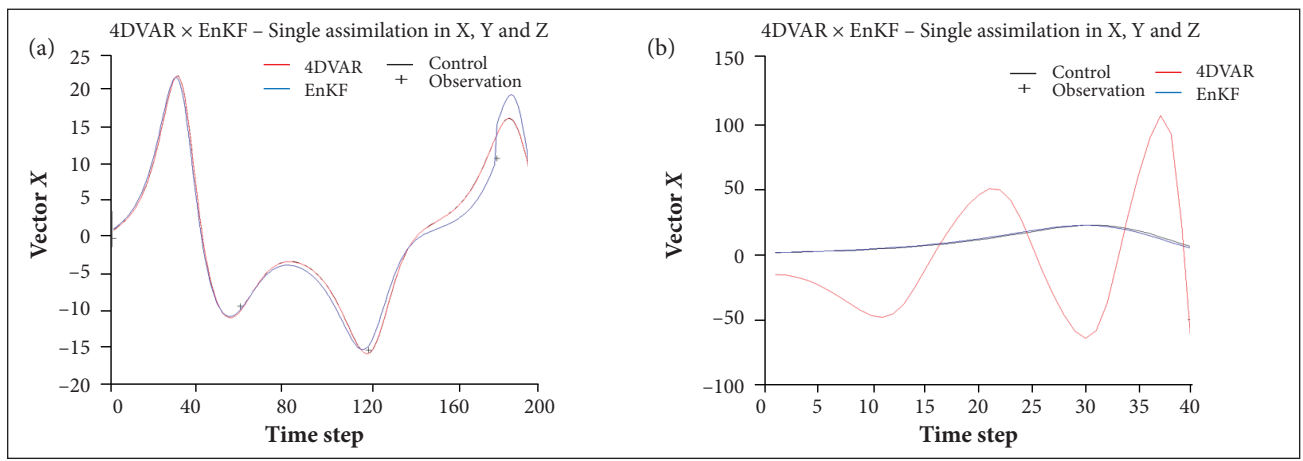

**Figure 3.** EnKF and 4DVAR ingest a single observation at the $180^{th}$ time step.

## FINAL COMMENTS

In this study, an EnKF and 4DVAR were available in a Data Assimilation context in chaotic regime. Both methodologies were examined and are effective to track the Control for 10, 20 and 40% of error at the IC. Considering 10% of noise at the IC, the results show a perfect fitting between assimilation curves and the Control. These results are still quite good for 20% of noise, but there is a disagreement between the "truth" and the estimation, especially at the end of integration period for EnKF, due the chaotic nature of the system. Considering 40% of noise at the IC, both EnKF and 4DVAR fail.

Regarding an experiment in which it is explored the assimilation over-determined by lack of observation, results show that EnKF needs more frequent observations, and the 4DVAR demands for observation in more than 1 variable of the Lorenz system.

Despite the limitations, EnKF and 4DVAR are state-of-the-art techniques in Data Assimilation implemented in numerical weather forecasting.

## ACKNOWLEDGEMENTS

The authors would like to thank the anonymous reviewers for their valuable comments and suggestions to improve the quality of this paper.

## AUTHOR'S CONTRIBUTION

Conceptualization, Harter FP and Corrêa CS; Methodology, Harter FP; Investigation, Harter FP and Corrêa CS; Writing – Original Draft, Harter FP; Writing – Review & Editing, Harter FP and Corrêa CS.

## REFERENCES

Anderson JL (2001) An ensemble adjustment filter for data assimilation. Mon Wea Rev 129(12):2884-2903. doi: 10.1175/1520-0493(2001)129<2884:AEAKFF>2.0.CO;2

Bishop C, Etherton B, Majumdar S (2001) Adaptive sampling with the ensemble transform Kalman filter part I: The theoretical aspects. Mon Wea Rev 129(3):420-436. doi: 10.1175/1520-0493(2001)129<0420:ASWTET>2.0.CO;2

Burgers G, Leeuwen PJV, Evensen G (1998) Analysis scheme in the Ensemble Kalman Filter. Mon Wea Rev 126(6):1719-1724. doi: 10.1175/1520-0493(1998)126<1719:ASITEK>2.0.CO;2

Courtier P, Andersson E, Heckley W, Vasiljevic D, Hamrud M, Hollingsworth A, Rabier F, Fisher M, Pailleux J (1998) The European Centre for Medium-Range Weather Forecasting (ECMWF) implementation of three dimensional variational assimilation (3D-Var). I: Formulation. Quart J Roy Met Soc 124(550):1783-1807. doi: 10.1002/qj.49712455002

Courtier P, Talagrand O (1998) Variational assimilation of meteorological observations with the adjoint vorticity equation. II: Numerical results. Quart J Roy Met Soc 113(478):1329-1347. doi: 10.1002/qj.49711347813

Courtier P, Thépaut JN, Hollingsworth A (1994) A strategy for operational implementation of 4D-VAR, using an incremental approach. Quart J Roy Met Soc 120(519):1367-1387. doi: 10.1002/qj.49712051912

Evensen G (1994) Sequential data assimilation with a nonlinear quasi-geostrophic model using Monte Carlo methods to forecast errors statistics. J Geophys Res 99(C5):10143-10162. doi: 10.1029/94JC00572

Kalnay E (2004) Atmospheric modeling, data assimilation and predictability. Cambridge: Cambridge University Press.

Lorenz EN (1963) Deterministic nonperiodic flow. J Atmos Sci 20:130-141. doi: 10.1175/1520-0469(1963)020<0130:DNF>2.0.CO;2

Miller R, Guil M, Gauthiez F (1994) Advanced data assimilation in strongly nonlinear dynamical systems. J Atmos Sci 51:1037-1056. doi: 10.1175/1520-0469(1994)051<1037:ADAISN>2.0.CO;2

Ott E, Hunt BR, Szunyogh I, Corazza M, Kalnay E, Patil DJ, Yorke JA, Zimin AV, Kostelich EJ (2002) Exploiting local low dimensionality of the atmospheric dynamics for efficient ensemble Kalman filtering; [accessed 2017 July 18]. https://pdfs.semanticscholar.org/1ea0/8d6362e2b02a47bbfca917d8789136c7b3af.pdf?_ga=2.94420245.867004641.1500426177-773654667.1496360440

Ott E, Hunt BR, Szunyogh I, Zimin AV, Kostelich EJ, Corazza M, Kalnay E, Patil DJ, Yorke JA (2004) A local ensemble Kalman filter

for atmospheric data assimilation. Tellus 56(5)::415-428. doi: 10.1111/j.1600-0870.2004.00076.x

Parrish DF, Derber JC (1992) The National Meteorological Center's spectral statistical-interpolation analysis system. Mon Wea Rev 120(8):1747-1763. doi: 10.1175/1520-0493(1992)120<1747: TNMCSS>2.0.CO;2

Rabier F, Courtier P (1992) Four-dimensional assimilation in the presence of baroclinic instability. Quart J Roy Met Soc 118(506):649-672. doi: 10.1002/qj.49711850604

Rabier F, Jarvinen H, Klinker E, Mahfouf JF, Simmons A (2000) The ECMWF operational implementation of four-dimensional variational physics. Quart J Roy Met Soc 126(564):1143-1170. doi: 10.1002/qj.49712656415

Saltzman B (1962) Finite amplitude free convection as an initial value problem. J Atmos Sci 19(4):329-341. doi: 10.1175/1520-0469(1962)019<0329:FAFCAA>2.0.CO;2

Sasaki Y (1970) Some basic formalisms in numerical variational analysis. Mon Wea Rev 98(12):875-883. doi: 10.1175/1520-0493(1970)098<0875:SBFINV>2.3.CO;2

Tippet MK, Anderson JL, Bishop CH, Hamill TM, Whitaker JS (2003) Ensemble square-root filters. Mon Wea Rev 131(7):1485-1490. doi: 10.1175/1520-0493(2003)131<1485:ESRF>2.0.CO;2

Whitaker JS, Hamill TM (2002) Ensemble data assimilation without perturbed observations. Mon Wea Rev 130(7):1913-1923. doi: 10.1175/1520-0493(2002)130<1913:EDAWPO>2.0.CO;2